\documentclass[aps,pra,preprint,groupedaddress]{revtex4-2}
\usepackage{graphicx,caption,subcaption}
\newcommand{\eqref}[1]{(\ref{#1})}%refer to equations
\renewcommand{\varphi}{\wp} % define \varphi to be \wp (Weierstrass rho)
\begin{document}

%\preprint{}

\title{Comment on "Doubly periodic solutions of the focusing nonlinear Schr\"odinger equation: Recurrence,
period doubling, and amplification outside the conventional modulation - instability band"}


\author{Hans Werner Sch\"urmann}
\email[]{hwschuer@uos.de}
\affiliation{Department of Physics\\ University of Osnabr\"uck, Germany}

\author{Valery Serov}
\email[]{vserov@cc.oulu.fi}
\affiliation{Department of Mathematical Sciences\\ University of Oulu, Finland,\\
Moscow Centre of Fundamental and Applied Mathematics-\\ - Lomonosov Moscow State University, Russia}


%\date{}

\begin{abstract}
In their interesting article (Physical Review A, Vol. 101, 023843 (2020)) Conforti et al. present doubly periodic (elliptic) solutions of the nonlinear Schr\"odinger  equation, based on an earlier article by Akhmediev et.al. (Theoretical and Mathematical Physics, Vol. 72, 809 (1987)). We present some notes with respect to correctness, completeness, and representation of the solutions obtained.
%It seems doubtful that the solution ansatz proposed is appropriate to solve the nonlinear Schr\"odinger equation. We outline arguments for this claim.

\end{abstract}

%\keywords{}

\maketitle

%\section{}
%\subsection{}
%\subsubsection{}

\section{Introduction}

In a recent article [1]
the authors present analytical solutions of the focusing Cubic Nonlinear Schr\"odinger Equation (CNLSE), following a seminal article by 
Akhmediev, Eleonskii, and Kulagin [2]. It seems that there are two flaws in [1] and [2]:\\
First, $Q(t, z)$, according to Eq.(6) in [1], is not a solution of the associated Eq.(15) in [2] (since in [1] only the final form of the solutions are presented, we refer to [2], if necessary). Second, consistency of Eq.(6) in [1] with Eq.(5) in [2] has not been checked neither in [1] nor in [2]. Furthermore, solution $Q(t, z)$ in [1] (see Fig.1) does not satisfy Eq.(5) in [2].

To specify our criticisms, in the following Section, we first shortly recapitulate some forms following the line presented in [2] and, second derive the correct solution $Q(t, z)$ of Eq.(6) in [2] (to be used, for the consistency check with Eq.(5) in [2]). In Section III we present simplifications and derive constraints for real, bounded solutions. In Section IV we return to the problem of consistency of Eq.(6) in [1] with Eq.(5) in [2], taking into account results of Section II. The Comment concludes with a summary. 

\section{Explicit Elliptic solutions}

In [1] the CNLSE in the "standard nonlinear fiber optics notation" 
\begin{equation}
i \Psi_z(t, z) + \Psi_{tt}(t, z) +a \Psi(t, z)|\Psi(t, z)|^2 = 0.
\end{equation}
is considered, where $z$ is the distance along the fiber, and $t$ is the (retarded) time. The solution ansatz 
\begin{equation}
\Psi(t, z) = (f(t, z) + i d(z)) e^{i\phi(z)},\quad f, d, \phi \in \mathbf{R},
\end{equation}
substituted into (1) leads to the system (of imaginary and real parts of CNLSE) (see Eqs.(4) and (5) in [2])
$$
(a) \quad 
f_z(t, z)=d(z)(\phi(z)-a(d^2(z)+f^2(t, z))),
$$
\begin{equation}
(b)\quad 
f_{tt}(t, z)=d_z(z)+(\phi_z(z)-ad^2(z))f(t, z)-af^3(t, z).
\end{equation}
As outlined in [2], Eq.(3b) can be integrated once so that the Frobenius compatibility condition $f_{zt}=f_{tz}$ can be applied, leading finally to a reduction of Eq.(3b) to three ordinary differential equations
\begin{equation}
(h_{z}(z))^2 = \alpha_1 h^4(z) + 4\beta_1 h^3(z) + 6\gamma_1 h^2(z) + 4\delta_1 h(z) + \epsilon_1 =: R_1(h),
\end{equation}
with
$$
\alpha_1 = - 16a^2,\quad \beta_1 = 4ac_1,\quad \gamma_1 = - \frac{1}{3}(2c_1^2 + 8ac_2),\quad \delta_1 = 2c_3,\quad \epsilon_1 = 0,
$$
\begin{equation}
(f_t(t, z))^2 = \alpha_2 f^4(t, z) + 4\beta_2 f^3(t, z) + 6\gamma_2 f^2(t, z) + 4\delta_2 f(t, z) + \epsilon_2 =: R_2(f, z),
\end{equation}
with
$$
\alpha_2 = - \frac{a}{2},\quad \beta_2 = 0,\quad \gamma_2 = \frac{1}{6}(c_1 - 3h(z)),\quad \delta_2 = \frac{h_z(z)}{4\sqrt{h(z)}},\quad \epsilon_2 = 2c_2 + \frac{3}{2}ah^2(z) - c_1h(z),
$$
\begin{equation}
\phi_z(z) = -2ah(z)+c_1
\end{equation}
with $d^2(z) = h(z)$ and integration constants $c_1, c_2, c_3$.

Eqs.(4), (6), (5) correspond to Eqs.(13), (14), (15) in [2], respectively. 
Obviously, the solutions of (4) and (5) are elliptic functions. Using a known (but seemingly not well known) formula due to
Weierstrass (see [3b], [4, Eq.(6)]) the solution of (4) reads 
$$
h(z)=
$$
\begin{equation}
\frac{4\wp(z)(h_0\wp+\beta_1h_0^2+2\gamma_1h_0+\delta_1)+2\wp_z(z)\sqrt{R_1(h_0)}+
h_0^2(2\alpha_1\delta_1-2\beta_1\gamma_1)+h_0(4\beta_1\delta_1-5\gamma_1^2)-2\gamma_1\delta_1}{(2\wp(z)-\gamma_1-2\beta_1h_0-\alpha_1h_0^2)^2-\frac{\alpha_1}{2}R_1(h_0)},
\end{equation}
where $\wp(z)=\wp(z; g_{2z}, g_{3z})$ denotes Weierstrass' function and
\begin{equation}
g_{2z}=3\gamma_1^2-4\beta_1\gamma_1,
\end{equation}
\begin{equation}
g_{3z}=-\gamma_1^3+2\beta_1\gamma_1\delta_1-\alpha_1\delta_1^2,
\end{equation}
\begin{equation}
\Delta_z=(3\gamma_1^2-4\beta_1\delta_1)^3-27(\gamma_1^3-2\beta_1\gamma_1\delta_1+\alpha_1\delta_1^2)^2
\end{equation}
are the invariants and discriminant of Weierstrass' function, respectively. In seeking real, bounded, nonnegative $h(z)$,
the boundary value $h_0=h(0)$ must be chosen appropriately (see below). The period $L_z$ of $h(z)$ is equal to the real period $2\omega$ of $\wp(z; g_{2z}, g_{3z})$ [5, Fig.18.1]
\begin{equation}
L_z=2\omega(g_{2z}, g_{3z}).
\end{equation}
With (7), integration of $\phi_z(z)$ according to (6)
leads to:
$$
\phi(z) = (c_1-2a)z + \phi(0) - \frac{a}{\sqrt{\alpha_1}}\log\frac{\gamma_1-2\wp(z)+h_0(2\beta_1+\alpha_1h_0)+\sqrt{\alpha_1R_1(h_0)}}{\gamma_1-2\wp(z)+h_0(2\beta_1+\alpha_1h_0)-\sqrt{\alpha_1R_1(h_0)}}+
$$
\begin{equation}
\frac{2a(r_3-r_1)(r_2-r_3)}{(r_3-r_4)\wp'(v_1)}\left(\log\frac{\sigma(z-v_1)}{\sigma(z+v_1)}+2z\zeta(v_1)\right)+\frac{2a(r_1-r_4)(r_2-r_4)}{(r_3-r_4)\wp'(v_2)}\left(\log\frac{\sigma(z-v_2)}{\sigma(z+v_2)}+2z\zeta(v_2)\right),
\end{equation}
where $\sigma, \zeta$ are Weierstrass-sigma and Weierstrass-zeta functions, respectively, with
$$
v_1 = \wp^{-1}(r_3; g_{2z}, g_{3z}),\quad v_2 = \wp^{-1}(r_4; g_{2z}, g_{3z}),
$$
$$
r_1 = -\frac{1}{2h_0}\left(\delta_1+2\gamma_1h_0+\beta_1h_0^2-\sqrt{\beta_1^2h_0^4+(6\beta_1\gamma_1-2\alpha_1\delta_1)h_0^3+(9\gamma_1^2-2\beta_1\delta_1)h_0^2+6\gamma_1\delta_1h_0+\delta_1^2} \right), 
$$
$$
r_2 = -\frac{1}{2h_0}\left(\delta_1+2\gamma_1h_0+\beta_1h_0^2+\sqrt{\beta_1^2h_0^4+(6\beta_1\gamma_1-2\alpha_1\delta_1)h_0^3+(9\gamma_1^2-2\beta_1\delta_1)h_0^2+6\gamma_1\delta_1h_0+\delta_1^2} \right), 
$$
\begin{equation}
r_3 = \frac{1}{2}\left(\gamma_1-2\beta_1h_0+\alpha_1h_0^2-\sqrt{\alpha_1R_1(h_0)}\right),\quad r_4 = \frac{1}{2}\left(\gamma_1+2\beta_1h_0+\alpha_1h_0^2-\sqrt{\alpha_1R_1(h_0)}\right).
\end{equation}

Using the same method as applied to (4), the solution of (5) reads 
$$
f(t, z)=
$$
\begin{equation}
\frac{-2\gamma_2\delta_2 - (5\gamma_2^2-\alpha_2\epsilon_2)f_0 + 2\alpha_2\delta_2 f_0^2 + 4\wp(t)(\delta_2+2\gamma_2f_0+\wp(t)f_0) + 
2\wp_t(t)\sqrt{R_2(f_0, z)}}{(2\wp(t)-\gamma_2-\alpha_2f_0^2)^2-\alpha_2R_2(f_0, z)},
\end{equation}
where $\wp(t)=\wp(t; g_{2t}, g_{3t})$ with
\begin{equation}
g_{2t}=\alpha_2\epsilon_2 + 3\gamma_2^2,
\end{equation}
\begin{equation}
g_{3t}=\alpha_2\gamma_2\epsilon_2  - \alpha_2\delta_2^2 - \gamma_2^3,
\end{equation}
\begin{equation}
\Delta_t=(\alpha_2\epsilon_2 + 3\gamma_2^2)^3 - 27(\alpha_2\gamma_2\epsilon_2  - \alpha_2\delta_2^2 - \gamma_2^3)^2.
\end{equation}
As for $h_0$, $z-$independent initial value $f_0=f(0, 0)$ must be chosen such that $f(t, z)$ is real and bounded. The period $L_t$ of $f(t, z)$ is equal to the real period $2\omega$ of $\wp(t; g_{2t}, g_{3t})$
\begin{equation}
L_t=2\omega(g_{2t}, g_{3t}; z).
\end{equation}
It depends on $z$ via $h(z)$. We note that the singularities of $\wp$ in (7), (12), (14) do not induce singularities of $h(z), \phi(z)$, and $f(t, z)$.

%At this point, some remarks are appropriate. Disregarding the different representation of $Q(t, z)$ according to Eq.(6) in [1] (or (24) in [2]) in terms of Jacobi functions, and $f(t, z)$, in terms of Weierstrass' functions, we note that $f(t, z)$ is different from $Q(t, z)$ because the modulus $k_q$ of $Q(t, z)$ does not depend on $z$. $Q(t, z)$ is the solution of Eq.(15) in [2] (with disregards to different notations of the variables), which is equation (4) above. The coefficients in both equations are dependent on $z$ and on $t$. Thus it is not clear why the $z-$dependence  drops out from the modulus of Eq.(6) in [1]. Moreover, the validity of solution (6) in [1] is not justified: If we assume that $Q(t, z)$ according to (6) in [1] is a solution of the corresponding equation (15) in [2] then the modulus $m_q$ in (6) is not correct. If we assume that $Q(t, z)$ with modulus $m_q$ is correct then $Q(t, z)$ is not a solution of equation (15) in [2].

Comparing formulae (7), (12) and (14) with the corresponding formulae in [1], [2], differences are obvious:\\
(i) Apart from $a$ and an integration constant $\phi(0)$ in (12) the family $\Psi(t, z)$ according to (2) depends on three integration constants $c_1, c_2, c_3$, boundary value $h_0 = h(0)$, and initial
value $f_0 = f (0, 0)$ at the boundary $z = 0$. It should be noted that $ h_0$ and $f_0$ are essential to find the constraints for real, bounded (”physical”) $h(z), f (t, z)$, as outlined below.\\
(ii) Solutions $\delta(z)$ and $Q(t, z)$ (see Eqs.(4), (6), (19), (20) in [1]) are expressed in terms of Jacobi elliptic functions (with $\alpha_i$ as the roots of the fourth degree polynomial, Eq.(13) in [2]). Despite the equivalence of Jacobi and Weierstrass elliptic functions, it is not a matter
of preference to use one of both for representations of $h(z)$ and $f(t, z)$ (see [4], summary).
Varying parameters (e.g., W, H, D in [2]) are leading to various $\alpha_i$ and hence (in general) to different Jacobi functions as in [1] (Eqs.(4) and (19)). Our representation of $\delta(z)$ and
$Q(t, z)$ as $h(z)$ and $f(t, z)$ according to (7) and (14) makes this discrimination unnecessary,
since (7) and (14) are valid, independently on the sign of $\Delta_z$ . Since $f(t, z)$ is triggered by $h(z)$, variable modulus is the ”normal” case. Thus, compared with (7), (14), it seems (at
least) inexpedient to use Jacobi functions for evaluation.\\
(iii) Solution (4) in [1] is a particular case of solution (7) ($h_0 = 0$ and three positive roots of $R_1(h) = 0$).  Additionally, due to $\delta(0) = 0$ according to (4) in [1], function value $Q(0, 0) = Q_D = \sqrt{\alpha_1} - \sqrt{\alpha_2} - \sqrt{\alpha_3}$ (see (6) in [1]) is special compared with
the range of $f_0$ defined by constraint $R_2(f_0, z) \ge 0$ in (14). Needless to say that free (or free in a certain domain) parameters are important for matching with experimental data; it
seems that fixed $\delta(0)$ and $Q(0, 0)$ are unnecessarily restrictive.\\
(iv) Finally, $f(t, z)$ is different from $Q(t, z)$ because
the modulus $k_q$ of $Q(t, z)$ does not depend on $z$. $Q(t, z)$ is the solution of Eq.(15) in [2]
(with disregards to different notations of the variables), which is equation (5) above. The
coefficients in both equations are dependent on $z$ and (on $t$ in [2]). Thus it is not clear why the
$z-$dependence drops out from the modulus of Eq.(6) in [1] (Eq.(24) in [2]). Moreover, the validity of solution
(6) in [1] is not justified: If we assume that $Q(t, z)$ according to (6) in [1] is a solution of the
corresponding equation (15) in [2] then the modulus $k_q$ in (6) is not correct. If we assume
that $Q(t, z)$ with modulus $k_q$ is correct then $Q(t, z)$ is not a solution of equation (15) in [2].

\section{Simplifications and Constraints }

%Solution (4) in [1] (or (22) in [2]) is a particular case of generic solution (7) ($h_0=0$ and three positive roots of $R_1(h)=0$). As a consequence, the doubly-periodic background is a particular one in the set of elliptic function backgrounds defined by (7) and (14). 
The above expressions for $h(z)$ and $\phi(z)$ can be simplified by considering the graphs $\{h_z^2(z), h(z)\}$, denoted as phase diagrams (PDs). It is well known that a phase diagram analysis is a useful tool for studying solutions of the nonlinear Schr\"odinger equation (see [6], [7]). Physical solutions must satisfy the phase diagram conditions (PDCs) [4] with roots of (4) denoted as PDC-roots [4, Section IV]. If $h_0$ is a (simple) PDC-root, we have $R_1(h_0)=0$, so that  $h(z)$ can be simplified as
\begin{equation}
h(z)=\frac{4\wp(z)(h_0\wp+\beta_1h_0^2+2\gamma_1h_0+\delta_1)+
h_0^2(2\alpha_1\delta_1-2\beta_1\gamma_1)+h_0(4\beta_1\delta_1-5\gamma_1^2)-2\gamma_1\delta_1}{(2\wp(z)-\gamma_1-2\beta_1h_0-\alpha_1h_0^2)^2}.
\end{equation}
In most of the 19 PDs of Fig.2 in [4] $h_0=0$ is a simple PDC-root. Thus, in these cases, we obtain 
\begin{equation}
h(z)=\frac{\delta_1}{\wp(z; g_{2z}, g_{3z})-\frac{\gamma_1}{2}}.
\end{equation}
Function $\phi(z)$ can be simplified correspondingly. With (12) we get (by taking the limit $h_0\to 0$ in (12), (13))
\begin{equation}
\phi(z) = (c_1-2a)z + \phi(0) 
-\frac{2a}{\wp'(v_1)}\left(\log\frac{\sigma(z-v_1)}{\sigma(z+v_1)}+2z\zeta(v_1)\right),\quad v_1=\wp^{-1}(\frac{\gamma_1}{2}; g_{2z}, g_{3z}).
\end{equation}
Equation (20) describes all physical solutions that correspond to PDs with simple PDC-root $h_0=0$. The associated allowed parameters $c_1, c_2, c_3, a$ can be found easily using the fact that the discriminant of $R_1(h_0)$ is equal to $\Delta_z$ (apart from a positive factor), by applying by the Cartesian sign-rule to the first quadrant of the PD only (due to $h(z)\ge 0$). The behaviour of $h(z)$ can be classified by (8)-(10): If $\Delta_z \ne 0$ or ($\Delta_z=0, g_{2z}>0, g_{3z}>0$), $h(z)$ is periodic, if $\Delta_z=0, g_{2z}\ge 0, g_{3z}\le 0, h(z)$ is solitary-like.

Similar considerations to simplify $f(t, z)$ and to specify $f(t, z)$ by $\Delta_t$ (as outlined before for $h(z)$ by $\Delta_z$) must take into account that $R_2(f_0, z), \Delta_t, g_{2t}, g_{3t}$ are depending on $z$. Unlike $R_1(h_0)=0$ with $h_0=0$, the simplifying condition $R_2(f_0, z)=0$ is not satisfied by $f_0=0$ in general, independently on $z$. $R_2(f_0, z)\ge 0$ defines a region in the $\{f_0, z\}-$plane with boundary $R_2(f_0, z)=0$ so that the phase diagrams are $3-$dimensional. -- Within the scope of this Comment we disregard a phase diagram analysis of $R_2(f_0, z)$. 

With respect to constraints for physical $h(z), f(t, z)$, we first consider $h(z)$, since physical $h(z)$ is necessary for physical $f(t, z)$.
Obviously (see Eq.(7)), if, first, the constraint  
\begin{equation}
R_1(h_0)>0,\quad h_0>0,
\end{equation} 
holds, the denominator in Eq.(7) is positive (due to $\alpha_1<0$). Function $h(z)$ is bounded, and thus it is physical if the numerator $N_1(h_0, z)$ in (7) is non-negative. If, second, $h_0=0$, function $h(z)$ is given by (20). The lower bound of $\wp(z; g_{2z}, g_{3z})$ is $e_1$, where $e_1$ is the largest positive root of $4j^3 - g_{2z}j - g_{3z}=0$ [8]. 
Thus, if 
\begin{equation}
e_1>\frac{\gamma_1}{2}, \quad \delta_1>0,
\end{equation}
then $h(z)$ is physical.
If, third, $R_1(h_0)=0, h_0>0$, function $h(z)$ is physical if the denominator in Eq.(7) is not equal  to zero, and the numerator $N_1(h_0, z)$ is non-negative. Thus, 
\begin{equation}
e_1>\frac{1}{2}(\gamma_1 + 2\beta_1h_0 + \alpha_1h_0^2), \quad N_1(z, h_0)\ge 0,
\end{equation}
in this case. 

Second, we consider $f(t, z)$ according to Eq.(14). The constraint, valid for certain domains in the  $\{f_0, z\}-$plane
\begin{equation}
R_2(f_0, z)\ge 0
\end{equation}
is necessary for real $f(t, z)$. 

For focusing fiber medium ($a>0$) we have $\alpha_2<0$, so that $f(t, z)$ is physical, if $R_2(f_0, z)>0$. If $R_2(f_0, z)=0$ in Eq.(14), then 
\begin{equation}
\tilde{e}_1(g_{2t}, g_{3t}; z)>\frac{1}{2}(\gamma_2(z)+\alpha_2f_0^2)
\end{equation}
must hold, where $\tilde{e}_1(g_{2t}, g_{3t}; z)$ is the largest positive root of $4j^3-g_{2t}j-g_{3t}=0$.

For defocusing material ($a<0$), subject to (25),
\begin{equation}
\tilde{e}_1(g_{2t}, g_{3t}; z)>\frac{1}{2}\left(\gamma_2(z)+\alpha_2f_0^2\pm\sqrt{\alpha_2R_2(f_0, z)}\right)
\end{equation}
is necessary for $f(t, z)$ to be physical. 

Subject to the constraints (22)-(24) for physical $h(z)$ (valid for certain parameters $h_0, c_1, c_2, c_3, a$), the constraints (25)-(27) define certain regions in the $\{f_0, z\}-$plane as mentioned above, meaning in particular that $f(t, z)$ is not physical for all $z$ (in general). Unlike the above restrictions for  $h_0$, that depend on the parameters $c_1, c_2, c_3, a$ only, the restrictions of the initial value $f_0$ depend (in particular) on the various $h(z)$ and hence on the restricted possible boundary value $h_0=h(0)$. In this case, knowledge of the various physical $h(z)$ is essential for obtaining the various elliptic doubly-periodic backgrounds $f(t, z)$ by evaluation of the constraints (25)-(27) (see example in Appendix). The compact representation of $h(z)$ and $f(t, z)$ by (7) and (14), respectively, opens the possibility to study the modulation of $f(t, z)$ via $h(z)$ by varying parameters $h_0, c_1, c_2, c_3, a, f_0$ and in dependence on $z$. Physical $h(z)$ presupposed, for evaluation the constraints (25)-(27) play a pivotal role, since they define the admissible domains for $f_0$ and $z$. The structure of these domains can be rather different for different parameters. With parameters as chosen in Appendix, the domain is arc-wise connected with unrestricted $z$ (see Fig.3). Other parameters are leading to, e.g., sub-domains occurring periodically w.r.t. $z$, or to only one simply connected domain (if, e.g., $h(z)$ is solitary-like). The boundaries of the admissible domains are important for studying instabilities of $f(t, z)$ (as PDC-roots are important for instabilities of $h(z)$ and thus of $f(t, z)$). If $f_0$ is selected such that, for certain $z_0$, $\{f_0, z_0\}$ is on the boundary, function $f(t, z)$ is unstable at $z=z_0$ (see Fig.4c). A different kind of instability is related to the behaviour of $h(z)$ due to the various phase diagrams. Phase diagram according to Fig.2(i) in [4] can serve as an example. If $h_0$ is varied from $h_0=0$ to the simple PDC-root, function $h(z)$ is switching from dark solitary behaviour to a constant ($h=h_0$) and then to bright solitary behaviour. This instability of $h(z)$ modulates $f(t, z)$, thus leading to an instability of $f(t, z)$.  

\section{On the adequacy of ansatz (2)}

As mentioned in the Introduction, consistency of $Q(t, z)$ (according to (6) in [1]) with Eq.(5) in [2] has not been checked. Thus it leads to the question whether $f(t, z)$ according to (14) is consistent with Eq.(3a). With $\phi_z$ from Eq.(6), Eq.(3a) can be written as 
\begin{equation}
f_z(t, z) = \sqrt{h(z)}(c_1-a(3h(z)+f^2(t, z))).
\end{equation}
This (Riccati-type) equation must be satisfied identically with physical $f(t, z), h(z)$ substituted. With parameters of the function background, presented in Appendix, numerical evaluation shows that (28) is not satisfied, leading to the problem, whether or not, subject to the PDC and the constraints, parameters $c_1, c_2, c_3, a, h_0, f_0$ exist, so that (28) is valid. The solution of this problem is crucial for some articles published in the past. Recently, with reference to [1] and [2], Conte published an interesting article [9] that could open a possibility to solve the consistency problem of system (3): Instead of solving the $t-$elliptic Eq. (5) with $z-$dependent coefficients (as outlined above), in [9], a solution of the $z-$Riccati equation (3a) (with $t-$dependent coefficients) has been presented. Solution (32) in [9] must be compared with solution (14), in order to get the additional constraints of the parameters. It seems that this is an intricate exercise. We consider it outside the scope of a Comment on [1] and [2].  
%-- For particular parameters $a=-1, c_1=-1, c_2=-\frac{c_1^2}{4a}, c_3=0, h_0=\frac{c_1}{a}, f_0=0$ Eq.(28) can be satisfied \underline{asymptotically} for $t, z\to \pm \infty$. For this case, $h(z)$ is bright solitary like with $h(z)\to 0, z\to \pm \infty$, (unbounded) $f(t, z)$ tends to zero for $t, z\to \pm \infty$, hence $f_z(t, z)\to 0$ in (28). Another possibility to solve the consistency problem of system (3) is to find a solution of (28), e.g. by using (20) for simplicity, and compare it with (14), in order to get restrictions for the parameters. -- Due to difficulty to solve (28), we disregard the possibility within the scope of this Comment: The adequacy of ansatz (2) remains an unsolved problem.

\section{Conclusion}

Induced by doubts in the correctness of $Q(t, z)$ in [1], we derived explicit solutions of system (3) expressed in terms of $h(z), \phi(z), f(t, z)$ together with constraints for reality and boundedness of these functions. Though the functions are well-defined (if the PDCs are taken into account), the numerical example of the elliptic function background shows that Eq.(3a) is not satisfied in general, leading to the unsolved problem whether Eq.(5) in [2] (or [28] above) can be satisfied by certain parameters (consistent with the PDC) or not. If not, ansatz (2) is not appropriate to solve the CNLSE. The main points of our criticism are:\\
 -- $Q(t, z)$ according to Eq.(6) in [1] (Eq.(24) in [2]) is flawed.\\
 -- The representation of $Q(t, z)$ in terms of Jacobi functions is not effective for numerical evaluation.\\
 -- Consistency of Eq.(5) in system (4), (5) with $Q(t, z)$ has not been checked. Equation (6) in [1] does not satisfy Eq.(5) in [2] (or [28]) above).

\section*{Appendix: EXAMPLE OF AN ELLIPTIC FUNCTION BACKGROUND}

Due to $\alpha_1<0$ in Eq.(4) and for simplicity we assume $\Delta_z>0, c_3>0$ and three changes of sign in (4). Hence we obtain the phase diagram according category (a) in Fig.2 in [4], that defines a particular family of two solutions. With parameters $c_1=-2, c_2=0.4, c_3=0.13, a=-1$, we consider the solution $h(z)$ with $0\le h_0\le 0.08$ (see PD, Fig.1). Choosing $h_0=0$, $h(z)$, according to Eq.(7), is given by (20).
It is positive and bounded with period $L_z=2.85$ (see (11)) as depicted in Fig.2.
To determine the range of possible $\{f_0, z\}$ for physical $f(z, t)$, constraint (25) together with constraint (27) must be evaluated. The result is shown in Fig.3 (appropriate $\{f_0, z\}$ in the red marked region). Obviously, $f_0=0$ is consistent with (25) and (27) (for "all" $z$), so that $f(z, t)$ and the doubly-periodic background $|\Psi|^2=f^2+h$ are physical (see Figs.4a, 4b). For $f_0=0.8$, unstable $|\Psi|^2$ is shown in Fig.4c.  The $z-$dependent real $t-$period $L_t=2\omega(z)$ of $f(z, t)$ according to (14) is depicted in Fig.5. The $z-$period of $L_t$ is equal to the real period of $h(z)$, given by (18). 
Evaluation of $\phi(z)$ according to Eq.(21) is straightforward with result shown in Fig.6.

%FIG1
\begin{figure}
\includegraphics[width=10cm]{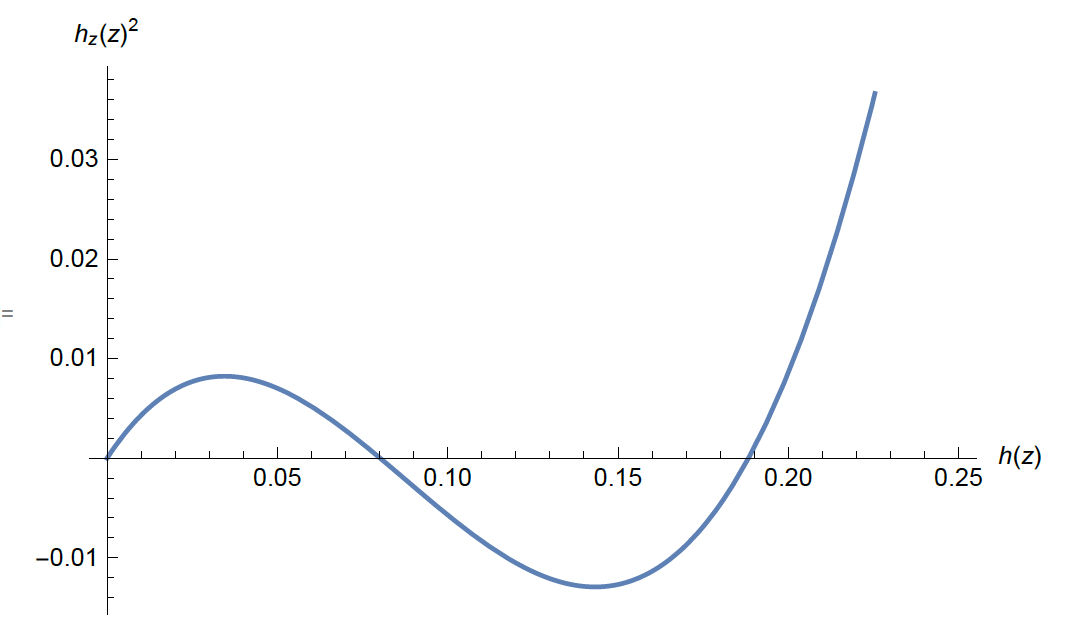}
\caption{Part of phase diagram $\{h_z^2(z), h(z)\}$. Parameters $c_1=-2; c_2=0.4; c_3=0.13; a=-1$.}
\end{figure}

\begin{figure}
\includegraphics[width=10cm]{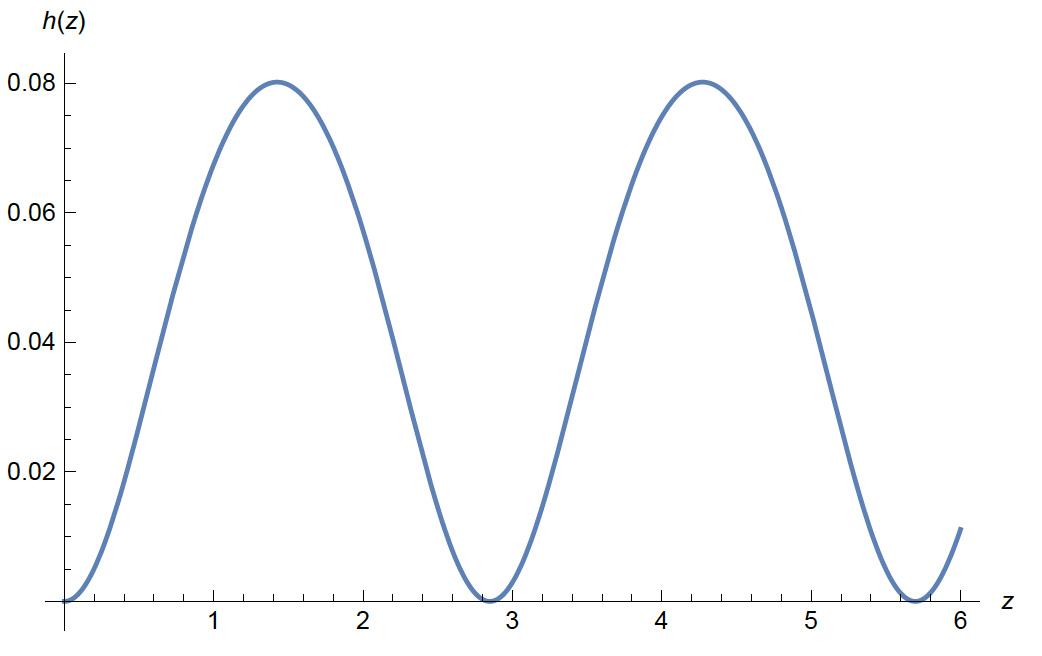}
\caption{Function $h(z)$ for $h_0=0$. Parameters as for Fig.1.}
\end{figure}

\begin{figure}

\begin{subfigure}[b]{0.33\textwidth}
         \centering
         \includegraphics[width=\textwidth]{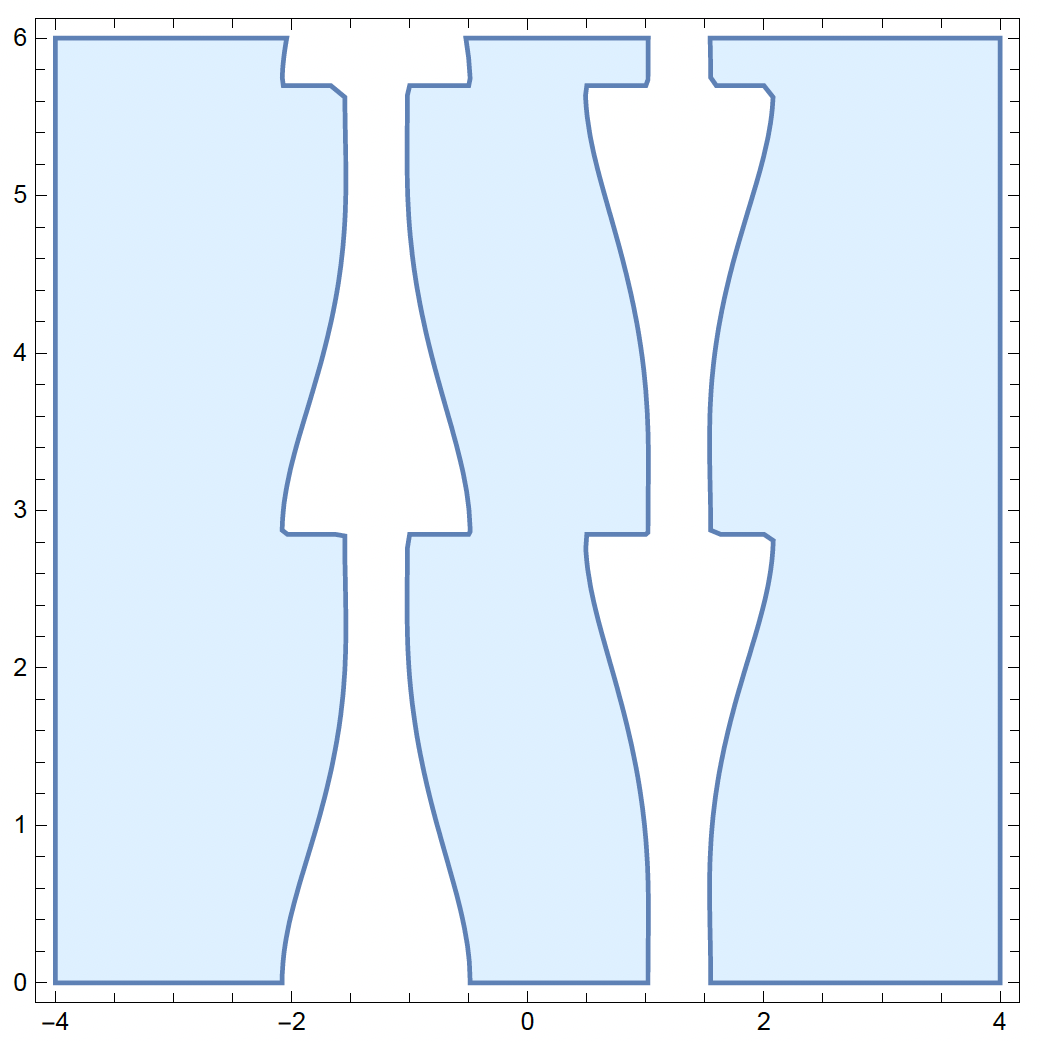}
         \caption{}
         \label{fig3a}
\end{subfigure}
\begin{subfigure}[b]{0.33\textwidth}
         \centering
         \includegraphics[width=\textwidth]{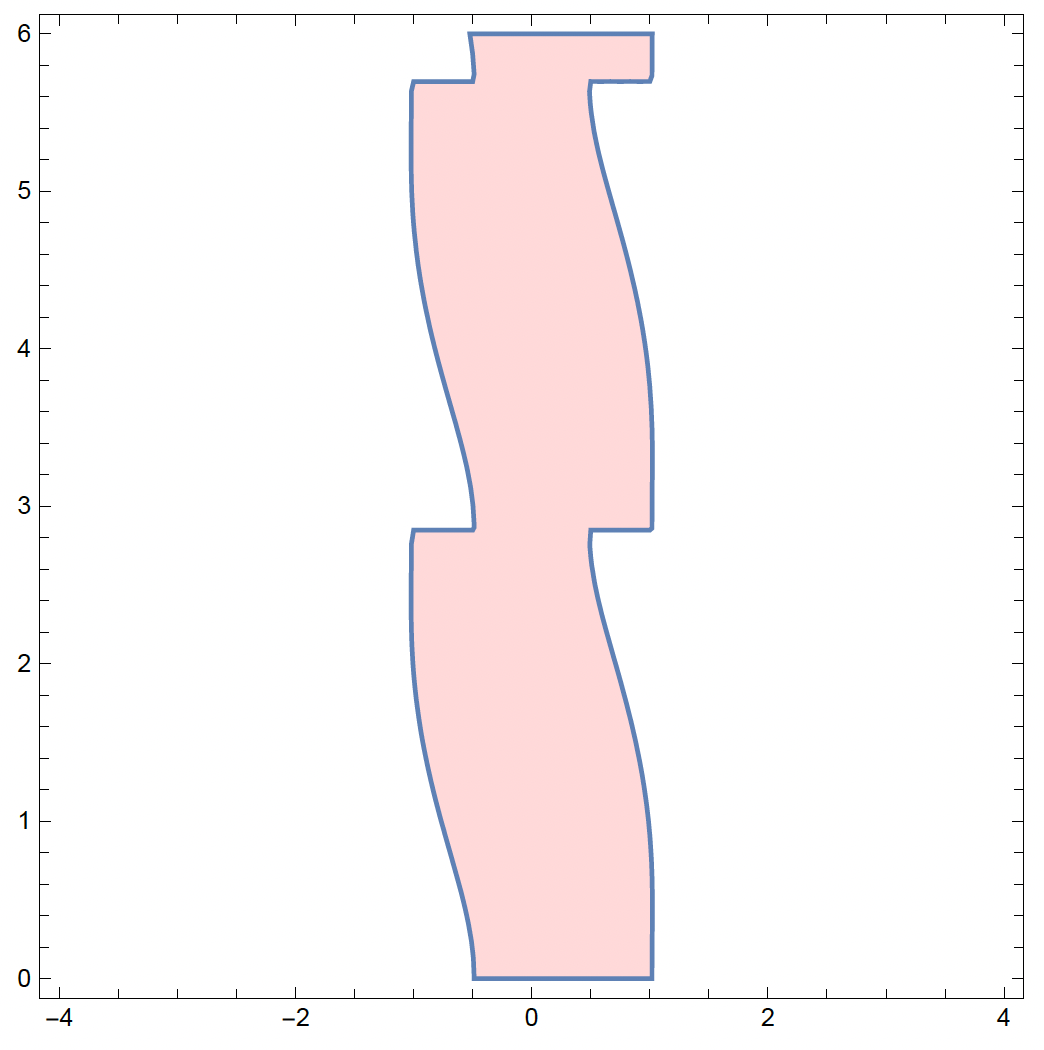}
         \caption{}
         \label{fig3b}
\end{subfigure}
\begin{subfigure}[b]{0.33\textwidth}
         \centering
         \includegraphics[width=\textwidth]{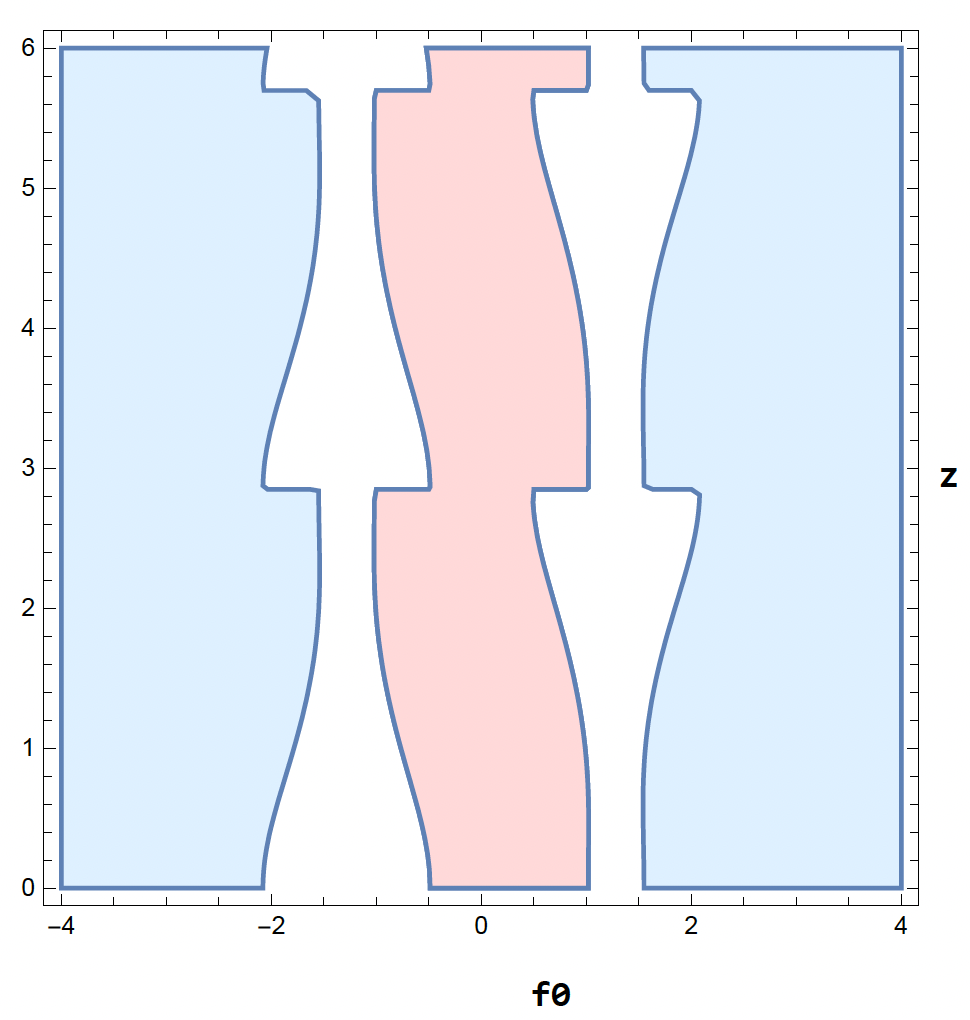}
         \caption{}
         \label{fig3c}
\end{subfigure}

\caption{Regionplots of constraint (25), (a), of constraint (27), (b), and the intersection of both. Allowed $\{f_0, z\}$ in the red region (c). Parameters as for Fig.1.}
\end{figure}

\begin{figure}

\begin{subfigure}[b]{0.35\textwidth}
         \centering
         \includegraphics[width=\textwidth]{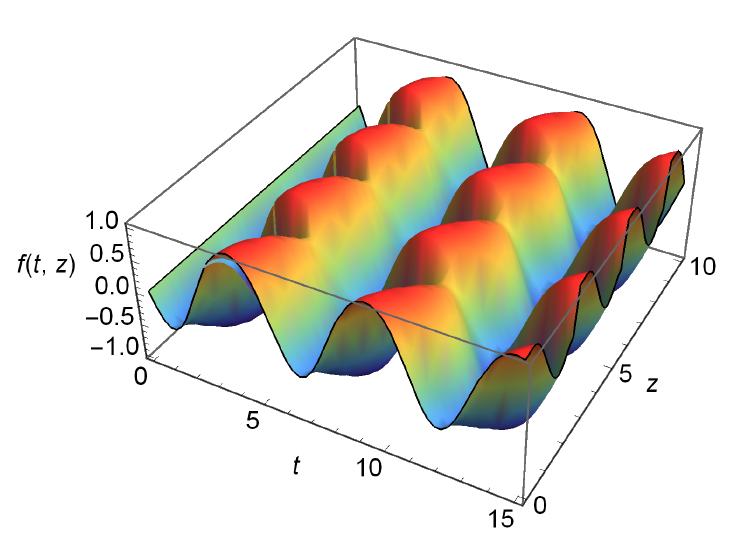}
         \caption{}
         \label{fig4a}
\end{subfigure}
\begin{subfigure}[b]{0.35\textwidth}
         \centering
         \includegraphics[width=\textwidth]{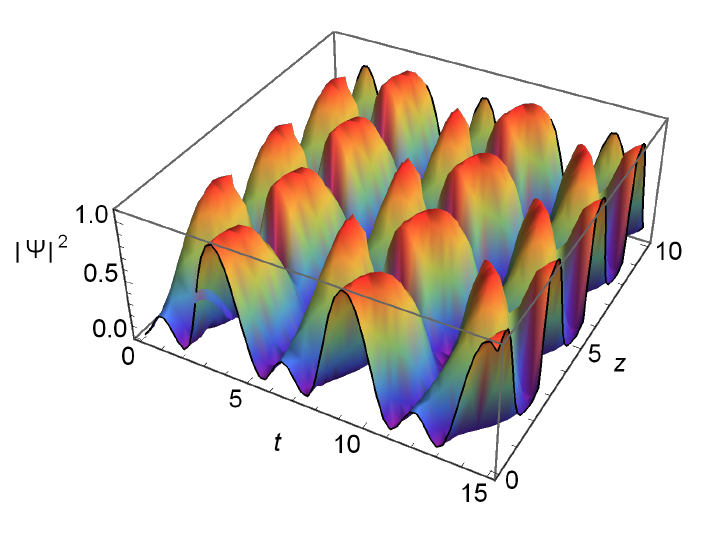}
         \caption{}
         \label{fig4b}
\end{subfigure}
\begin{subfigure}[b]{0.35\textwidth}
         \centering
         \includegraphics[width=\textwidth]{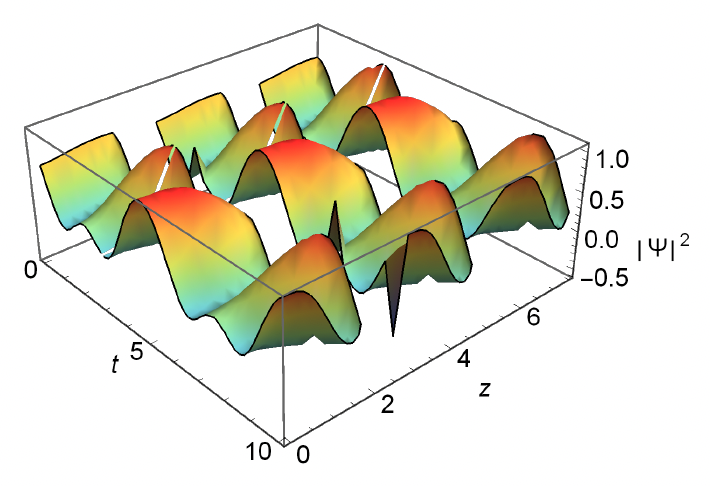}
         \caption{}
         \label{fig4c}
\end{subfigure}

\caption{Solution surfaces $f(t, z)$, (a); solution surface $|\Psi|^2$  with $f_0=0$, (b); solution surface $|\Psi|^2$ with $f_0=0.8$, (c). Parameters as for Fig.1.}
\end{figure}

\begin{figure}
\includegraphics[width=10cm]{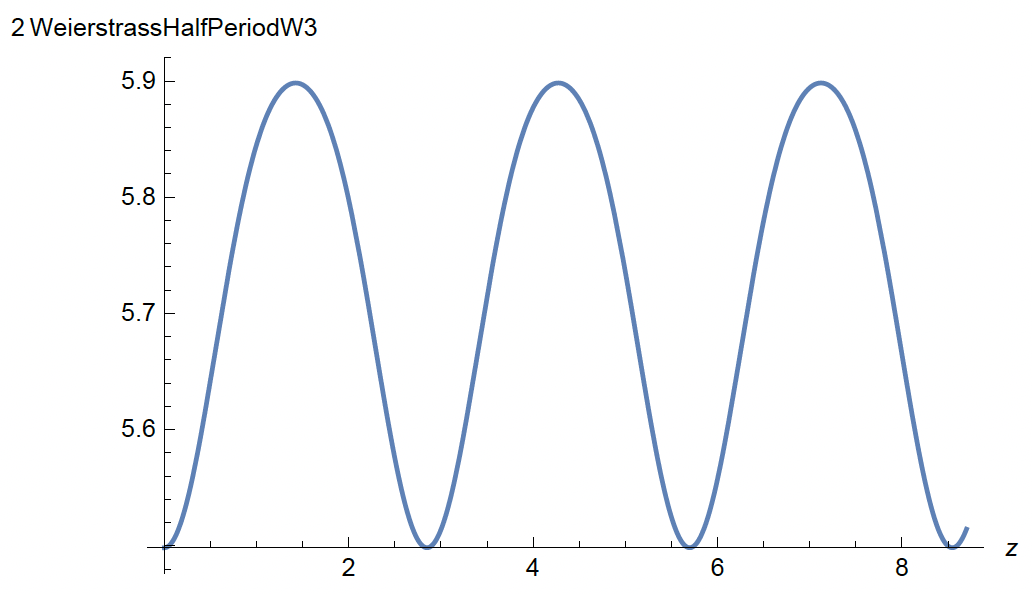}
\caption{Period $L_t=2\omega(g_{2t}(z), g_{3t}(z))$ of $f(t, z)$ as a function of $z$. Parameters as for Fig.1, $h_0=0, f_0=0$.}
\end{figure}

\begin{figure}
\includegraphics[width=10cm]{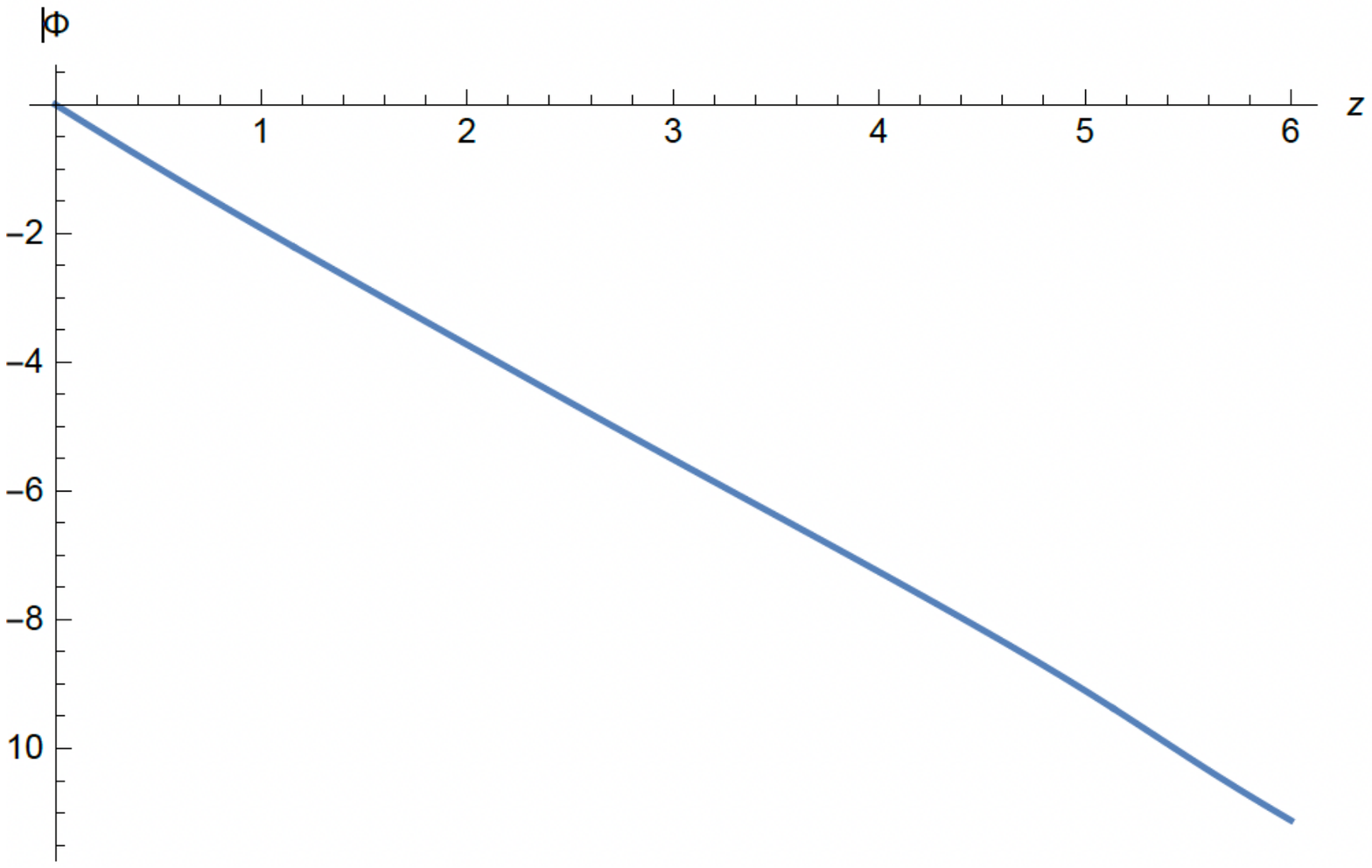}
\caption{Phase $\phi(z)$ according to Eq.(21). Parameters as for Fig.1.}
\end{figure}

\section*{References}

%Create the reference section using BibTeX:


\begin{thebibliography}{99}

%\bibitem{AkhA} N. Akhmediev and A. Ankiewicz, \emph{Solitons: Nonlinear Pulses and Beams}, (Chapmann and Hall, Boca Raton, 1997). 

\bibitem{CMKTA} M. Conforti, A. Mussot, A. Kudlinski, S. Trillo and N. Akhmediev, Physical Review A, Vol. 101, 023843  (2020). 

\bibitem{AkhEK} N. Akhmediev, V.M. Eleonskii and N.E. Kulagin, Theor. Math. Phys., Vol. 72, 809 (1987).

%\bibitem{C} R. Conte, Theor. Math. Phys., Vol. 209(1), 1366 (2021). 

%\bibitem{ChPW} J. Chen, D.E. Polinovsky and R.E. White, Physical Review E, Vol. 101, 052219 (2019). 


%\bibitem{WW} E.T. Whittaker and G.N. Watson, \emph{A Course of Modern Analysis}, p.454, Cambridge (1927). 

\bibitem{W}  (a) K. Weierstrass, \emph{Mathematische Werke V}, 14-16, (Johnson, New York, 1915); (b) E.T. Whittaker and G.N. Watson, \emph{A Course of Modern Analysis}, 454 (Cambridge University Press, Cambridge, 1927).

\bibitem{SS}  H.W. Sch\"urmann and V.S Serov, Physical Review A, Vol. 93, 063802 (2016).

%\bibitem{SSS}  H.W. Sch\"urmann, V.S Serov and Y.V. Shestopalov,  Physical Review E, Vol. 58, 1040 (1998).

\bibitem{Abr}  \emph{Handbook of Mathematical Functions}, edited by M. Abramowitz and I.A. Stegun (Dover, New York, 1968).

%\bibitem{rem} Assuming that solution (6) is correct, i.e., it satisfies Eq.(15) in [3] we obtain the contradiction either with (15) in [3] or with $k^2_q=\frac{\alpha_2-\alpha_1}{\alpha_3-\alpha_1}$ from (6) in [2].


\bibitem{GWG}  L. Gagnon and P. Winternitz, Phys. Rev. A, Vol. 39, 296 (1989); J. Phys. A, Vol. 21, 1493 (1988); Vol. 22, 469 (1989); L. Gagnon, B. Grammaticos, A. Ramani and P. Winternitz, ibid, Vol. 22, 499 (1989).

\bibitem{S}  H.W. Sch\"urmann, Phys. Rev. E, Vol. 54, 4312 (1996).


\bibitem{Tr} F. Tricomi, \emph{Elliptische Funktionen}, 52-62, (Akademische Verlagsgesellschaft, Leipzig, 1948).
%\bibitem{SS1} H.W. Sch\"urmann and V.S Serov, PIERS 2004 Pisa, Session 26a.

%\bibitem{VSN} G. Vanderhaegen, P. Szriftziger, C. Naveau et al., Optics Lett., 45, 3757 (2020).

\bibitem{C} R. Conte, Theor. Math. Phys., Vol. 209(1), 1366 (2021). 

\end{thebibliography}
\end{document}